\documentstyle[epsf]{mn}
\begin{document}
\title[Baade's Windows in the mid-infrared]
{ISOGAL Survey of Baade's Windows in the  Mid-infrared
}
\author[I.S. Glass, S. Ganesh et al.]
{
I.S. Glass$^1$, S. Ganesh$^{2,3}$, C. Alard$^4$, J.A.D.L. 
Blommaert$^5$, 
G. Gilmore$^6$,\cr 
T. Lloyd Evans$^1$, A. Omont$^2$, M. Schultheis$^2$ and G. Simon$^4$ \\
$^1$South African Astronomical Observatory, PO Box 9, Observatory 7935, 
South Africa\\ 
$^2$Institut d'Astrophysique de Paris, CNRS, 98bis Blvd Arago, Paris F75014,
France\\
$^3$on leave from Physical Research Laboratory, Navarangpura, Ahmedabad 
380009, India\\
$^4$DASGAL, UMR CNRS N$_\circ$ 335, Observatoire de Paris, 21 Avenue de 
l'Observatoire, Paris F75014, France\\
$^5$ISO Data Centre, Astrophysics Div., Space Science Dept.\ of ESA, 
Apartado 50727, Madrid E28080, Spain\\ 
$^6$Institute of Astronomy, The Observatories, Madingley Rd., Cambridge CB3
0HA, UK\\
}

\date{Sent to MNRAS 22 Jan 1999}
\maketitle

\begin{abstract}

The ISOGAL mid-infrared survey of areas close to the Galactic Plane 
aims to determine their stellar content and its possible bearing on 
the history of the Galaxy. The NGC\,6522 and Sgr I Baade's Windows 
of low obscuration towards the inner parts of the Bulge represent 
ideal places in which to calibrate and understand the ISOGAL 
colour-magnitude diagrams, which are more difficult to interpret in 
heavily reddened fields. 

The survey observations were made with the ISOCAM instrument of the ISO 
satellite. The filter bands chosen were LW2 ($\sim$7\,$\mu$m) and LW3 
($\sim$15\,$\mu$m).

The results presented here show that most of the detected objects are late
M-type giants on the AGB, with a cut-off for those earlier than M3--M4. 
The most luminous members of these two Bulge fields at 7\,$\mu$m are 
the Mira variables. However, it is evident that they represent the end of
a sequence of increasing 15\,$\mu$m dust emission which commences with M
giants of earlier sub-type.

In observations of late-type giants the ISOCAM 15\,$\mu$m band is mainly 
sensitive to the cool silicate or aluminate dust shells which overwhelm 
the photospheric emission. However, in ordinary M-giant stars, the 7\,$\mu$m 
band is not strongly affected by dust emission and may be influenced 
instead by absorption. The $\nu_2$ band of water at 6.25\,$\mu$m and 
the SiO fundamental at 7.9\,$\mu$m are likely contributors to this effect.  
Miras are more luminous at 7\,$\mu$m and have redder $K_0$--[7] colours than 
other M giants. Their [7]--[15] colours are consequently bluer than might 
be expected by extrapolating from warmer M-giants. 

A group of late M stars has been found which vary little or not at all
but have infrared colours typical of well-developed dust shells. Their
luminosities are similar to those of 200--300 day Miras but they have 
slightly redder [7]--[15] colours which form an extension of the ordinary 
M giant sequence. They may belong to the class of red semi-regular variables.

The Mira dust shells show a mid-infrared [7]--[15] colour-period 
relation.

In these two fields there is no component of high-luminosity late-type 
variables obscured at $K$ and shorter wavelengths such as is seen in 
the Magellanic Clouds. The upper limit of LPV periods in these fields 
remains at {\it ca} 700 days.

\end{abstract}

\begin{keywords}
Stars: AGB, Stars:variables:others, Galaxy:center, Galaxy: stellar content 

\end{keywords}

\section {Introduction}

The ISOGAL\footnote{This is paper no.\ 3 in a refereed journal based on 
data from the ISOGAL project.} (P\'{e}rault et al, 1996; Ojha, 
Omont and Simon, 1997, Omont et al, 1999 a,b,c) project has 
surveyed a number of fields at low galactic latitude 
in the intermediate infrared using the ISOCAM (Cesarsky et al., 1996) 
instrument of the ISO\footnote{ISO is an ESA project with instruments 
funded by ESA member states (especially the PI countries: France, 
Germany, the Netherlands and the United Kingdom)  and with the 
participation of ISAS and NASA.} satellite. Its aim is the better 
understanding of the stellar content of the inner Galaxy, mostly in 
very obscured regions. 
Normally, the filter bands chosen for the observations were LW2 
(5.5--8\,$\mu$m) and LW3 (12--18\,$\mu$m). The pixel size was 6 $\times$ 
6 arcsec$^2$. 

The dusty shells surrounding mass-losing late-type stars are expected to be 
particularly prevalent amongst the objects that should be detected. These 
stars will also, in general, be seen by the DENIS and 2-MASS surveys 
at IJK$_S$ and JHK$_S$ respectively. 

However, in the more obscured ISOGAL fields, there is little information 
available concerning individual stars except for those few that are 
among the most luminous OH/IR sources. For this reason, it was decided 
to include some of the relatively unobscured fields known as ``Baade's 
Windows", located in the inner Bulge, in the programme. These should 
serve as relatively well surveyed comparison areas. $A_V$ is thought to
average 1.78 $\pm$ 0.10 mag in these fields as a whole (see Glass 
et al, 1995)\footnote{An extinction map by Stanek (1996) indicates, 
however, that it is particularly low (about $A_V$ $\sim$ 1.45) in the 
part of the NGC\,6522 window covered by our observations. The $K_0$
magnitudes that we quote later in this work remain as published by the
original authors, who used $A_K$ $\sim$ 0.14 mag.}. In both cases 
it will be almost negligible in the LW2 and LW3 bands discussed here and 
the ISOGAL magnitudes have not been corrected for interstellar extinction. 

In particular, we have examined the field located around the globular 
cluster NGC\,6522 and that known as Sgr I. The late-type stellar contents 
(M-stars) of the first of these fields have been surveyed and classified 
by Blanco, McCarthy and Blanco (1984) (BMB) by objective prism, and a  
smaller portion has been examined similarly by Blanco (1986) to fainter 
magnitudes. $I$-band photometry is given for the BMB survey, together 
with an indication of variability, whereas $V$-band photometry is 
provided by Blanco (1986). The long-period variable star content of 
both fields has been surveyed by Lloyd Evans (1976; TLE), using $I$-band 
plates, and he summarizes previous work at visible wavelengths. 

Infrared ($JHKL$) studies of Lloyd Evans' variable stars in NGC\,6522 
and Sgr I were carried out by Glass and Feast (1982) in order to make use of 
their period-luminosity relation for determining the distance 
to the Galactic Centre. It was later pointed out by Feast (1985) that many 
of the IRAS sources in the Sgr I and NGC\,6522 windows could be identified 
with known variables. These were listed by Glass (1986) who showed that 
the remaining (unidentified) IRAS sources in Sgr I were very red at $JHKL$ 
and were also likely to be long-period variables. Following this work, the 
known long-period variables and IRAS sources in the Sgr I field were monitored 
by Glass et al (1995) and periods were confirmed or determined for all 
sources except one that was non-variable. The $K_0, (H-K)_0$ colour-magnitude 
diagram of the Sgr I field shows that the long-period variables are among 
the reddest and most luminous objects (Glass, 1993) at these wavelengths. 
The dispersion of the $K_0$, log\,$P$ relation in Sgr I is quite small
($\sim$ 0.35 mag; Glass et al., 1995). This implies that most of the 
AGB stars in the field are at a nearly uniform distance, allowing a simple
connection to be made between apparent and absolute magnitudes.

Photometry of some of the BMB (1984) stars has been obtained by Frogel 
and Whitford (1987). Deep near-infrared photometry of the NGC\,6522 
window is presented by Tiede, Frogel and Terndrup (1995). 
The DENIS results will form the subject of a separate paper.

\section{ISOGAL Observations}

The ISOGAL observations were made in the two filters mentioned, as
rasters covering squares of 15 $\times$ 15 arcmin$^2$ orientated in 
$\ell$,$b$. They were centered at $\ell$=+1.03$^{\circ}$, $b$=--3.83$^{\circ}$, 
which includes the globular cluster NGC\,6522 itself, and at 
$\ell$=+1.37$^{\circ}$, $b$=--2.63$^{\circ}$ in Sgr I. Each position on the
sky was observed for a total of 22s on average.

Two rasters of each field were made with the  LW2 filter and one with LW3
(Table 1). The first three digits of each identification indicate the ISO
revolution number ($\simeq$ day of flight). The second LW2 observation in
each case is almost simultaneous with that in LW3 (within $\sim$ 30 min).
The images are shown in Figs 1--4. The 6$''$ pixel field of view was
utilised in all cases.

\begin{table}
\caption{Journal of observations}
\begin{tabular}{lllll}
Field & Filter & Identification & Julian date \\
NGC\,6522 & LW2 & 47101493 & 2450509 \\
NGC\,6522 & LW2 & 84001115 & 2450877 \\
NGC\,6522 & LW3 & 84001116 & 2450877 \\
Sgr I     & LW2 & 47101494 & 2450509 \\
Sgr I     & LW2 & 83800913 & 2450875 \\
Sgr I     & LW3 & 83800914 & 2450875 \\
\end{tabular}
\end{table}

\begin{figure}
\epsfxsize=8cm
\epsffile[0 0 332 342]{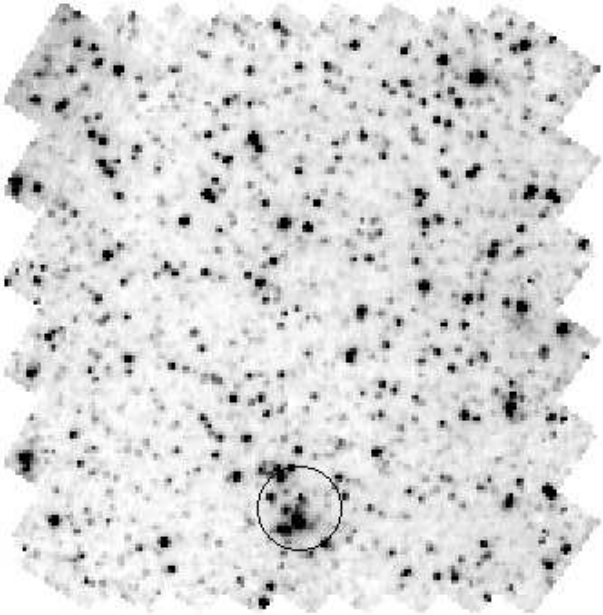}
\caption{ISOCAM 7\,$\mu$m (LW2) image of the field in the NGC\,6522 Baade
Window, covering 15 $\times$ 15 arcmin$^2$. Increasing $\ell$ is to the 
left and increasing $b$ is upwards. The NGC\,6522 cluster is visible 
near the middle of the bottom edge (Image 84001115). The circle has a radius
0.02$^{\circ}$.}
\end{figure}

\begin{figure}
\epsfxsize=8cm
\epsffile[0 0 332 342]{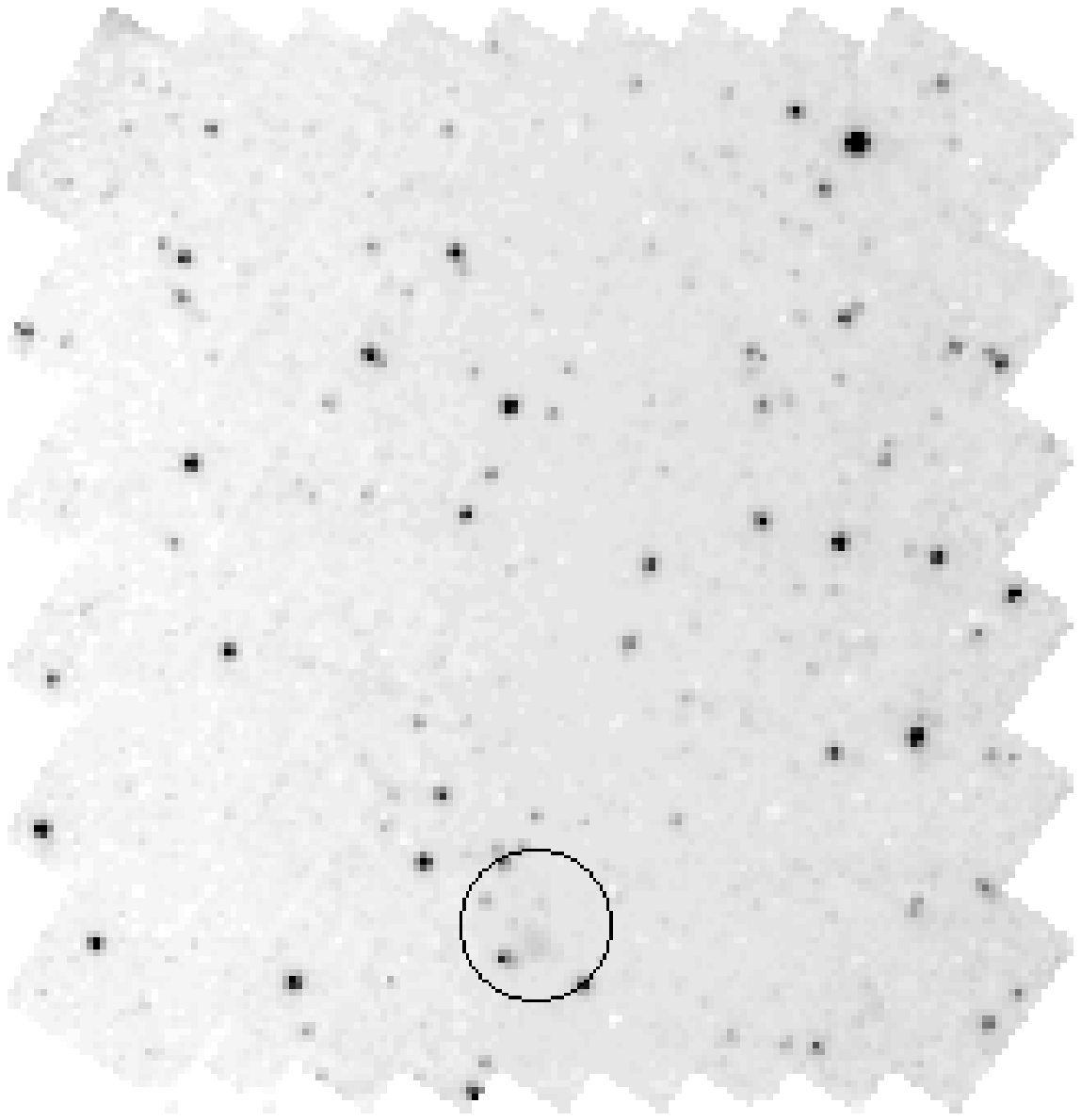}
\caption{ISOCAM 15\,$\mu$m (LW3) image of the field in the NGC\,6522 Baade
Window, covering 15 $\times$ 15 arcmin$^2$. Increasing $\ell$ is to the 
left and increasing $b$ is upwards (Image 84001116).}
\end{figure}

\begin{figure}
\epsfxsize=8cm
\epsffile[0 0 332 342]{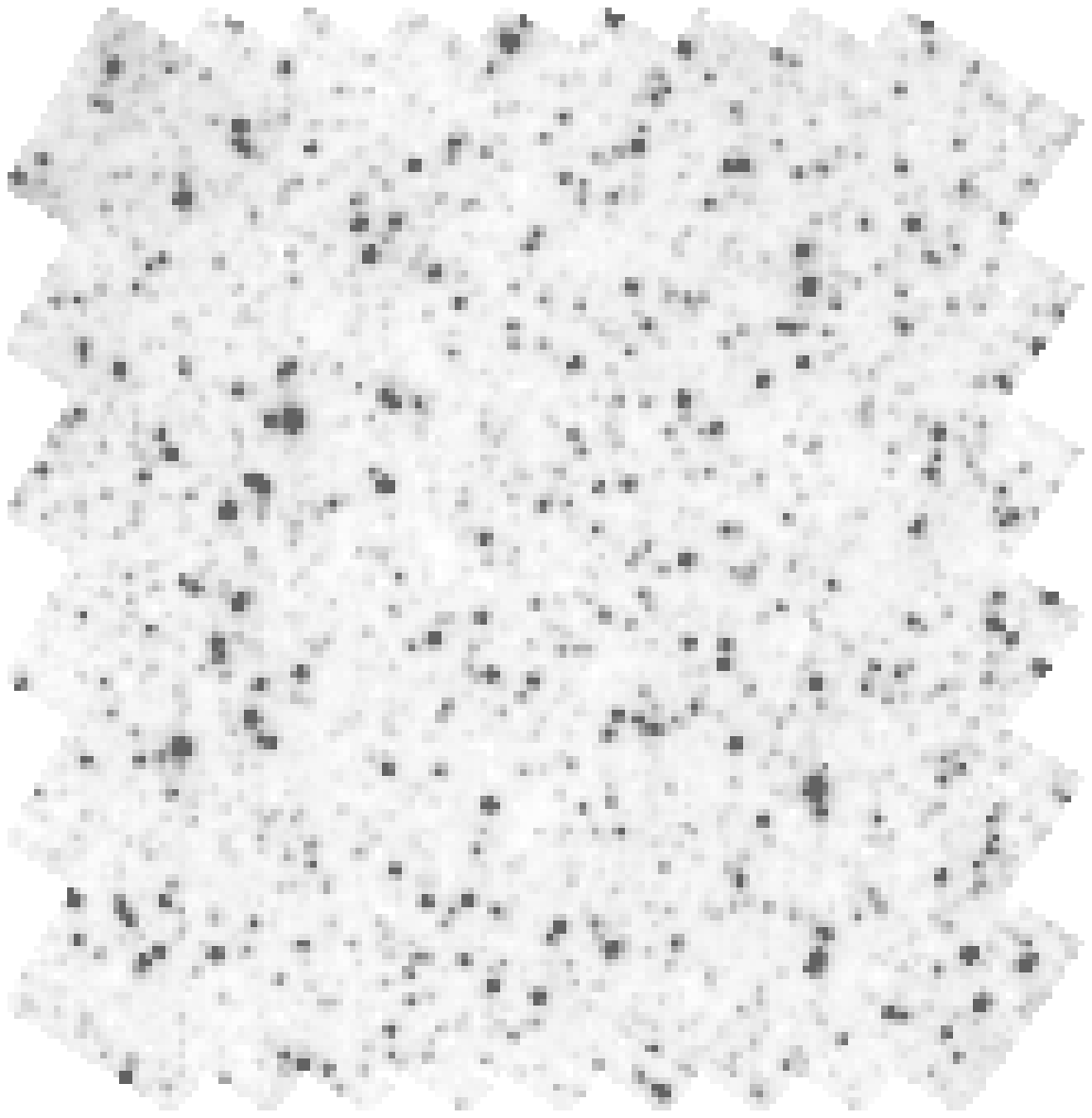}
\caption{ISOCAM 7\,$\mu$m (LW2) image of the field in the Sgr I Baade
Window, covering 15 $\times$ 15 arcmin$^2$. Increasing $\ell$ is to the 
left and increasing $b$ is upwards (Image 83800913).}
\end{figure}

\begin{figure}
\epsfxsize=8cm
\epsffile[0 0 332 342]{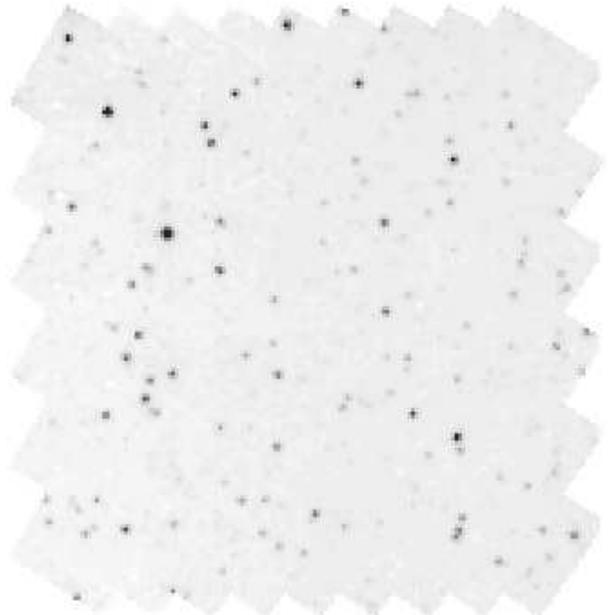}
\caption{ISOCAM 15\,$\mu$m (LW3) image of the field in the Sgr I Baade
Window, covering 15 $\times$ 15 arcmin$^2$. Increasing $\ell$ is to the 
left and increasing $b$ is upwards (Image 83800914).}
\end{figure}

Reduction of the science processed data (SPD) from version 6.32 of the OLP
(Off-line Processing) pipeline was carried out with the CIA\footnote{The 
ISOCAM data presented in this paper was analysed using `CIA', a joint 
development by the ESA Astrophysics Division and the ISOCAM Consortium.  
The ISOCAM Consortium is led by the ISOCAM PI, C. Cesarsky, Direction 
des Sciences de la Mati\`{e}re, C.E.A., France.} package (CIA version 3.0)
The data were first corrected for dark current, using the default method of
subtracting a `model' dark frame. Following this, the cosmic-ray hits were
removed from the data cube using the multi-resolution median method. At this
stage, two copies of the data cube were made and the individual copies were
treated by two different methods for simulating the time behaviour of the
pixels of the ISOCAM detectors: the `vision' method and the `IAS model 
transient correction', also known as the `inversion' method (Abergel et 
al., 1998). 
The vision method does not correct much for the transient behaviour 
but is useful in removing the memory remnants (electronic ghosts) 
of previously observed strong sources. From this stage onwards in 
the reduction procedure we thus had two sets of data, corresponding 
to the two methods of stabilization. For each data set the images 
at each raster pointing were averaged. The data were then flat-fielded 
with the flats generated from the raster observations themselves. 
Subsequent to the flat fielding, the individual images were 
mosaiced together after correcting for the field of view distortion using 
the `projection' method. 

The two individual rasters thus obtained were then in units of ADU/gain/sec. 
They were converted into mJy/pixel units within CIA. The conversion factors 
were:

\[
F({\rm mJy}) = (\rm ADU/gain/sec)/2.33  
\]
for LW2 and
\[
F({\rm mJy}) = (\rm ADU/gain/sec)/1.97.  
\]

\noindent for LW3 (Blommaert, 1998).

These are correct  for a $F_{\lambda}$ $\propto$ $\lambda^{-1}$ 
power-law spectrum at wavelengths 6.7\,$\mu$m and 14.3\,$\mu$m respectively. 

Source extraction was performed on each pair of reduced images using a point
spread function fitting routine (Alard et al., in preparation). The
vision-treated point sources were cross-identified with the
inversion-treated ones and the final catalogue of point sources was built
with the inversion photometry for those sources found in both vision- and
inversion-treated images. This procedure ensured that most false sources were
dropped while the better photometry of the inversion-treated images was
retained. 

Conversion to magnitudes was then carried out using the formulae
\[
[7] = 12.38 - 2.5 \; {\rm log} \; F_{\rm LW2}({\rm mJy}) 
\]
and
\[
[15] = 10.79 - 2.5 \; {\rm log} \; F_{\rm LW3}({\rm mJy})
\] 
where the zero point has been chosen to get zero magnitude for a Vega model 
flux at the respective wavelengths mentioned earlier. We have limited the
extracted catalogue to sources with fluxes greater  than 5mJy in both LW2
and LW3. This corresponds to [7] = 10.64 and [15] = 8.99.

The reliability of the ISOCAM data is affected at the faint end by crowding
and noise. These effects can be investigated by constructing histograms
of source counts vs mag. Figure 5 is a histogram of the LW3 observations in 
the NGC\, 6522 field. Also included is the expected distribution, based 
on averaged [7] magnitudes and a relation [15]$_{\rm expected}$ = 1.56 
$\times$ [7] -- 5.0 (This approximation is a line passing through the 
[15], [7]--[15] values of (9.0, 0.0) and (3.96, 1.8) in the 
colour-magnitude diagrams (Figures 8 and 9)).  It appears that the 
NGC\,6522 detections at 15\,$\mu$m are reliable to a depth 
of [15] $\sim$ 9.0 mag. 
Figure 6 is a similar diagram for the Sgr I 
field. 

\begin{figure}
\epsfxsize=8cm
\epsffile[28 322 539 786]{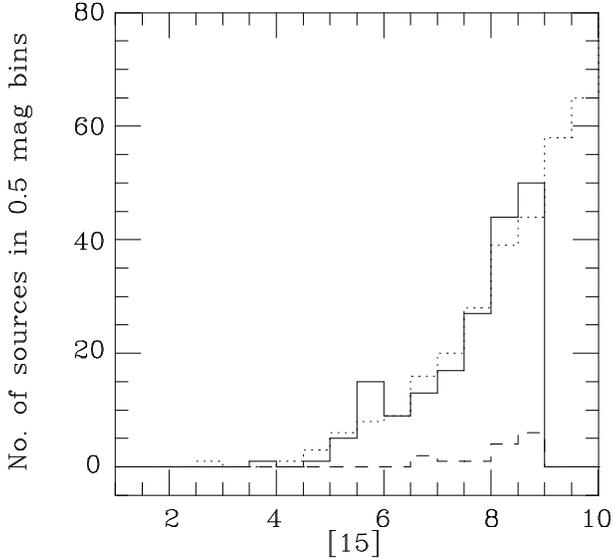}
\caption{Total observed (solid line) and expected (dotted line) distributions 
of all sources at 15\,$\mu$m (LW3) in the NGC\,6522 field, demonstrating 
that the LW3 detections are reliable to $\sim$ 9.0 mag (see text for 
explanation). The dashed histogram is the distribution of sources detected
only at 15\,$\mu$m. The latter may be spurious, especially near the faint
limit.}
\end{figure}  

\begin{figure}
\epsfxsize=8cm
\epsffile[28 322 539 786]{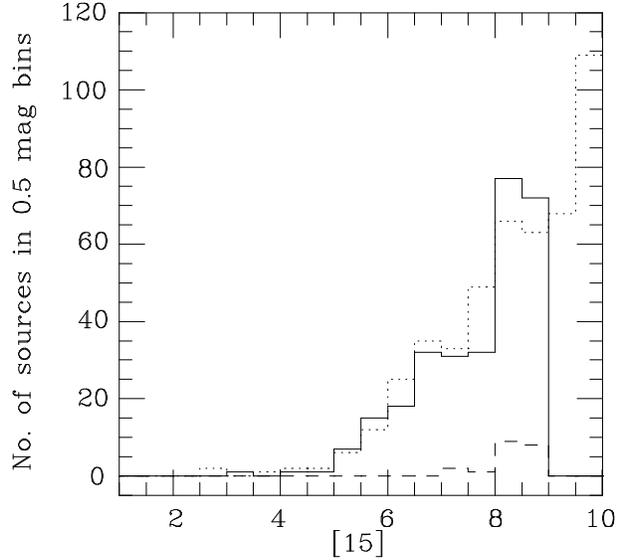}
\caption{Observed (solid line) and expected (dotted line) distributions for
the Sgr I field. See caption of Fig.\ 5 for further details.}
\end{figure}  

From analysis of repeated observations, the rms dispersion of the ISOGAL
photometry is estimated to be generally less than 0.2 magnitudes, 
except towards the fainter magnitude ranges, where it rises to 
$\sim$0.4 mag (Ganesh et al., in preparation). See also section 
3.3.3 of this paper. A small uncertainty remains in the absolute 
photometry due to the crowded nature of the ISOGAL fields. It is felt that
the behaviour of the inversion method of transient correction is not yet
completely understood and it is hoped that the uncertainty can be reduced 
in the near future.

It may be noted that, with $\sim$ 45 and $\sim$ 35 pixels per LW2 source, 
respectively, the density of sources in our two fields is close to the 
confusion limit in both.

The astrometry of the extracted sources was improved by using the newly 
available DENIS $K$ band observations of these fields.  A systematic 
shift of $0.8''$ (with a rms dispersion of 0.6$''$) in right 
ascension and $5.1''$ (0.7$''$) in declination was found for the NGC\,6522 
observation.  The corresponding numbers for the Sgr I observation were 
found to be $2.0''$ (0.9$''$) and $6.1''$ (0.8$''$) respectively.  The 
positions given in Tables 2 and 3 include the corrections.

\subsection{NGC\,6522}

Sources were extracted from the LW2 and LW3 images and cross-correlated to
form a working table. The total number detected was 497. Of these, 182 or 37\% were 
detected at 15\,$\mu$m.  The remainder were detected only in LW2, usually 
because they were too faint to be seen at 15\,$\mu$m. Sources were detected
in both LW2 exposures in 363 cases. Non-detections in LW2 are often
clearly due to blending with strong sources in their neighbourhoods. In
some cases the reliability of the detection is questionable and needs to be
confirmed when deeper data become available. Fifty-four sources were 
detected only in the first LW2 observation and 66 only in the second. 
These are for the most part close to the detection limit and their 
existence requires confirmation.

Some 14 sources were detected at 15\,$\mu$m only. Each was investigated 
visually in the three images and comments on individual sources were
included in the working table. 

Because of the small uncertainty in the zero-point of the photometry 
that has already been mentioned, we have decided to publish only those 
sources with [7] $\leq$ 7.5 for the time being (see table 2). The full 
table will be published when it is felt that the performance of the camera 
in crowded fields is better understood.  

The lack of a strong density enhancement in the neighbourhood of NGC\,6522 
indicates that few sources are likely to be members of the cluster itself.
Circles with radius 0.02 degree about the cluster centre are shown on figs.\ 1
and 2.

\setcounter{table}{1}

\begin{table*}

\begin{minipage}{18cm}

\caption{Bright sources detected by ISO in the NGC\,6522 field}

\begin{tabular}{lllllll}           

No. &  Name & [7]\#1 & [7]\#2 & [7]$_{avg}$ & [15] &  Cross-identifications and comments.\\ 
    1  & ISOGAL-PJ180228.2-300038  &    6.75  &          &    6.75  &          &    near raster edge - poor astrometry		\\ 
   12  & ISOGAL-PJ180234.8-295957  &    4.88  &    5.05  &    4.96  &    3.64  &    TLED9 400:d GF FW 				\\ 
   14  & ISOGAL-PJ180235.2-295855  &    6.46  &    6.53  &    6.50  &    5.78  &   *H 						\\ 
   26  & ISOGAL-PJ180239.4-295636  &    7.46  &          &    7.46  &    7.16  &  						\\ 
   41  & ISOGAL-PJ180242.9-300335  &          &    7.48  &    7.48  &    5.86  &  						\\ 
   46  & ISOGAL-PJ180243.5-300250  &    6.77  &    6.56  &    6.66  &    6.35  &  Unresolved? (See 49)				\\ 
   85  & ISOGAL-PJ180251.2-300013  &    7.59  &    7.39  &    7.49  &    6.61  &  BMB7  7					\\ 
  110  & ISOGAL-PJ180256.1-295533  &    6.77  &    6.64  &    6.71  &    5.77  &  *I						\\ 
  120  & ISOGAL-PJ180257.0-300417  &    6.23  &    6.76  &    6.50  &    5.54  &  *G BMB18 6					\\ 
  128  & ISOGAL-PJ180257.6-295124  &    6.95  &    6.65  &    6.80  &    6.32  &  						\\ 
  147  & ISOGAL-PJ180259.6-300253  &    6.92  &    7.13  &    7.03  &    5.28  &  *C BMB28 7 FW					\\ 
  162  & ISOGAL-PJ180301.1-300141  &    6.23  &    6.49  &    6.36  &    5.51  &  *B						\\ 
  170  & ISOGAL-PJ180302.3-295806  &    7.18  &    7.23  &    7.21  &    7.04  &  						\\ 
  180  & ISOGAL-PJ180303.4-295729  &    6.03  &    6.03  &    6.03  &    5.11  &  TLE 228 270d GF (3) BMB39 9 150		\\ 
  192  & ISOGAL-PJ180305.3-295515  &    6.84  &    6.87  &    6.86  &    5.52  &  *F BMB46 7 FW (Unresolved in LW2?)		\\ 
  205  & ISOGAL-PJ180306.3-295203  &    7.50  &    7.31  &    7.40  &    5.75  &  						\\ 
  218  & ISOGAL-PJ180307.8-300034  &    7.26  &    7.16  &    7.21  &    5.84  &  BMB52 9 FW					\\ 
  224  & ISOGAL-PJ180308.5-300524  &    6.39  &    6.25  &    6.32  &    4.82  &  TLE 403 335d GF BMB54 7 FW			\\ 
  235  & ISOGAL-PJ180309.8-295505  &    7.23  &    7.35  &    7.29  &    7.01  &  BMB61 7					\\ 
  245  & ISOGAL-PJ180311.5-295745  &    6.75  &    6.93  &    6.84  &    5.80  &  TLE 238 290: (2) BMB63 7 FW 			\\ 
  254  & ISOGAL-PJ180312.5-300428  &    6.99  &    7.10  &    7.04  &    5.98  &  BMB64 7					\\ 
  256  & ISOGAL-PJ180313.4-300055  &    7.42  &    7.33  &    7.38  &    6.35  &  BMB67 8					\\ 
  270  & ISOGAL-PJ180315.5-300737  &    7.42  &    7.25  &    7.33  &    6.57  &  BMB70 6.5  - at edge in LW2 (unresolved?)	\\ 
  299  & ISOGAL-PJ180318.4-295346  &    7.23  &    7.07  &    7.15  &    5.29  &  *D BMB86 9 FW					\\ 
  333  & ISOGAL-PJ180322.4-300255  &    7.43  &    7.43  &    7.43  &    7.08  &  B24 M6 BMB93 6 FW				\\ 
  382  & ISOGAL-PJ180328.4-295544  &    7.34  &    7.17  &    7.25  &    5.54  &  *E BMB119 9   (unresolved?)\\ 
  389  & ISOGAL-PJ180329.4-295939  &    7.36  &    7.40  &    7.38  &    5.99  &  B57 M7 BMB120 7 FW 				\\ 
  410  & ISOGAL-PJ180331.3-300100  &    6.77  &    6.65  &    6.71  &    6.33  &  BMB129 6.5  GC member?			\\ 
  432  & ISOGAL-PJ180334.1-295957  &    7.03  &    6.92  &    6.97  &    5.56  &  B81 M8 BMB142 8 FW 				\\ 
  441  & ISOGAL-PJ180335.7-300300  &    7.33  &    7.45  &    7.39  &    5.74  &  GC member? 					\\ 
  445  & ISOGAL-PJ180336.9-300147  &    7.12  &    7.02  &    7.07  &    6.03  &  BMB152 9 V   (Unresolved?) GC member?\\ 
  476  & ISOGAL-PJ180346.1-295911  &    6.91  &    7.08  &    7.00  &    5.21  &  *A B138 M7 BMB179 7 FW			\\	   
  491  & ISOGAL-PJ180350.9-295617  &    6.29  &    6.33  &    6.31  &    5.26  &  TLE136 270d (1) BMB194 6.5 FW			\\

\end{tabular}

\vspace{1mm}

Notes to tables 2 and 3:                                           
                  
The ``Name" column follows the IAU nomenclature. The letter ``P" denotes
provisional\\
1 Noted by Lloyd Evans (1976) as semiregular \\    
2 Noted by Lloyd Evans (1976) as small amplitude in $I$\\
3 =IRAS 17598-2957\\                               

Key:                                             
                                                 
TLE: LPVs in Lloyd Evans (1976)\\                 
GF: Miras with $JHKL$ photometry by Glass and Feast (1982)\\
\*A etc: stars selected for blink examination\\      
B: Stars classified by Blanco (1986)\\
BMB: stars classified by BMB (1984), followed by M sub-class\\ 
FW: $JHK$ photometry in Frogel \& Whitford (1987)\\

\end{minipage}                                                

\end{table*}

\subsection{Sgr I}

A total of 696 sources were detected. Of these, 287 or 41\% were detected at
15\,$\mu$m . Only 20 sources were detected solely at 15\,$\mu$m. 
At 7\,$\mu$m, 517 sources were detected twice. Eighty-eight sources were
only seen in the first LW2 exposure and 71 were only seen in the second.
As in the case of the NGC\,6522 field, only those sources with [7] $\leq$ 
7.5 are given in Table 3.

\setcounter{table}{2}

\begin{table*}

\begin{minipage}{180mm}

\caption{Bright sources detected in the Sgr I field by ISO}
\begin{tabular}{lllllll}
No & Name & [7]\#1 & [7]\#2 & [7]$_{avg}$ & [15] & Cross-identifications and comments\\
   32  & ISOGAL-PJ175836.4-290802  &    7.27  &    7.14  &    7.21  &    6.28  &  				\\ 
   45  & ISOGAL-PJ175839.1-290522  &    7.29  &    7.23  &    7.26  &    6.54  &  				\\ 
   61  & ISOGAL-PJ175841.8-290352  &    6.84  &    6.88  &    6.86  &    5.41  &   *D				\\ 
   65  & ISOGAL-PJ175842.0-290649  &    7.19  &    7.06  &    7.12  &    6.73  &   Unresolved?  	        \\ 
   69  & ISOGAL-PJ175842.6-291028  &          &    7.44  &    7.44  &    6.59  &   at edge of raster; outside [7]\#1 observation; unresolved?				\\ 
   78  & ISOGAL-PJ175843.9-290711  &    6.61  &    6.06  &    6.33  &    5.28  &   TLE65 Unresolved?            \\ 
  112  & ISOGAL-PJ175848.0-291002  &    7.32  &    7.34  &    7.33  &    7.26  &  				\\ 
  117  & ISOGAL-PJ175848.7-290743  &    7.41  &    7.55  &    7.48  &    6.58  &  				\\ 
  129  & ISOGAL-PJ175850.3-290454  &    7.43  &    7.52  &    7.47  &    6.57  &  				\\ 
  167  & ISOGAL-PJ175854.8-285831  &    6.93  &    6.76  &    6.84  &    6.94  &  				\\ 
  168  & ISOGAL-PJ175854.9-290117  &    7.40  &    7.53  &    7.46  &    7.76  &  				\\ 
  170  & ISOGAL-PJ175855.1-290627  &    6.95  &    7.03  &    6.99  &    5.53  &   *M   Unresolved?             \\ 
  180  & ISOGAL-PJ175855.9-285845  &    7.34  &    7.42  &    7.38  &    6.86  &  				\\ 
  187  & ISOGAL-PJ175856.4-290049  &    6.81  &    6.65  &    6.73  &    5.37  &   *F				\\ 
  189  & ISOGAL-PJ175856.6-290213  &    7.30  &    7.35  &    7.32  &    7.24  &  				\\ 
  200  & ISOGAL-PJ175857.4-291215  &    7.14  &    7.25  &    7.20  &    6.11  &  				\\ 
  206  & ISOGAL-PJ175857.8-290114  &    6.60  &    6.71  &    6.65  &    5.34  &   *J				\\ 
  247  & ISOGAL-PJ175901.1-285821  &    6.13  &    5.79  &    5.96  &    4.46  &   TLE79  Unresolved? 	        \\ 
  291  & ISOGAL-PJ175904.7-290744  &    7.01  &    7.03  &    7.02  &    5.65  &   *N				\\ 
  303  & ISOGAL-PJ175905.7-290235  &    6.99  &    7.17  &    7.08  &    6.01  &  				\\ 
  328  & ISOGAL-PJ175907.3-291024  &    7.17  &    7.31  &    7.24  &    6.68  &  				\\ 
  338  & ISOGAL-PJ175908.3-290855  &    7.53  &    7.38  &    7.46  &    6.40  &  				\\ 
  351  & ISOGAL-PJ175909.5-290903  &    7.34  &    7.40  &    7.37  &    7.27  &  				\\ 
  352  & ISOGAL-PJ175909.5-290825  &    7.52  &    7.45  &    7.48  &    7.31  &  				\\ 
  363  & ISOGAL-PJ175910.5-290129  &    4.66  &    5.08  &    4.87  &    3.37  &   TLE53 Unresolved? 	        \\ 
  365  & ISOGAL-PJ175910.6-290456  &    7.22  &    7.27  &    7.24  &    6.45  &  				\\ 
  375  & ISOGAL-PJ175911.1-290315  &    6.66  &    6.75  &    6.71  &    5.34  &   *E   Unresolved?             \\ 
  400  & ISOGAL-PJ175913.6-285812  &    7.37  &    7.34  &    7.36  &    7.10  &  				\\ 
  401  & ISOGAL-PJ175913.7-291113  &    5.03  &    4.97  &    5.00  &    4.77  &  				\\ 
  402  & ISOGAL-PJ175913.7-285852  &    7.04  &    6.88  &    6.96  &    5.75  &   TLE87			\\ 
  405  & ISOGAL-PJ175914.0-290950  &    7.24  &    7.05  &    7.14  &    5.58  &   *G				\\ 
  410  & ISOGAL-PJ175914.4-291335  &    7.24  &    7.18  &    7.21  &    6.80  &  				\\ 
  416  & ISOGAL-PJ175914.8-291129  &    7.46  &    7.35  &    7.40  &    6.41  &  				\\ 
  421  & ISOGAL-PJ175915.5-290133  &    6.79  &    6.82  &    6.80  &    6.07  &   appears unresolved.		\\ 
  434  & ISOGAL-PJ175916.8-290010  &    7.27  &    7.22  &    7.24  &    7.38  &  				\\ 
  446  & ISOGAL-PJ175917.7-285839  &    7.47  &    7.40  &    7.44  &          &  				\\ 
  452  & ISOGAL-PJ175918.2-291433  &    6.69  &    6.63  &    6.66  &    6.50  &  				\\ 
  453  & ISOGAL-PJ175918.2-290123  &    5.99  &    6.15  &    6.07  &    5.83  &  				\\ 
  455  & ISOGAL-PJ175918.5-290504  &    7.34  &    7.25  &    7.29  &    6.58  &  				\\ 
  457  & ISOGAL-PJ175918.5-290607  &    7.28  &    7.36  &    7.32  &    5.77  &  				\\ 
  478  & ISOGAL-PJ175920.7-291505  &    7.41  &    7.33  &    7.37  &    7.07  &  				\\ 
  488  & ISOGAL-PJ175922.0-291229  &    6.36  &    6.55  &    6.46  &    6.03  &  				\\ 
  511  & ISOGAL-PJ175923.4-290215  &    6.18  &    6.56  &    6.37  &    5.39  &   TLE54                        \\ 
  514  & ISOGAL-PJ175923.7-291236  &    6.76  &    6.91  &    6.83  &    5.64  &   TLE39			\\ 
  522  & ISOGAL-PJ175924.4-291358  &    7.24  &    7.24  &    7.24  &    6.21  &  				\\ 
  524  & ISOGAL-PJ175924.6-291236  &          &    7.47  &    7.47  &    6.15  &  				\\ 
  529  & ISOGAL-PJ175925.3-290336  &    7.05  &    7.22  &    7.13  &    5.70  &   *L				\\ 
  540  & ISOGAL-PJ175926.4-290707  &    7.15  &    6.90  &    7.03  &    5.84  &   		                \\ 
  541  & ISOGAL-PJ175926.5-290217  &    7.31  &    7.37  &    7.34  &    5.60  &   *H                   	\\ 
  550  & ISOGAL-PJ175927.4-290309  &    7.75  &    7.17  &    7.46  &    5.91  &  				\\ 
  556  & ISOGAL-PJ175927.9-290530  &    7.44  &    7.47  &    7.46  &    6.61  &  				\\ 
  576  & ISOGAL-PJ175929.7-290318  &    6.35  &    6.68  &    6.51  &    5.47  &   TLE55			\\ 
  581  & ISOGAL-PJ175930.7-290950  &    6.77  &    7.00  &    6.88  &    6.83  &  				\\ 
  585  & ISOGAL-PJ175931.1-290858  &    6.93  &    7.18  &    7.05  &    5.67  &   *K				\\ 
  601  & ISOGAL-PJ175932.8-290734  &    7.25  &    7.33  &    7.29  &    6.59  &  				\\ 
  609  & ISOGAL-PJ175934.0-290236  &    5.41  &    5.63  &    5.52  &    5.53  &   unresolved?                  \\ 
  613  & ISOGAL-PJ175934.4-290702  &    7.43  &    7.49  &    7.46  &    6.40  &  				\\ 
  626  & ISOGAL-PJ175936.1-290915  &    6.94  &    6.97  &    6.96  &    6.75  &  				\\ 
  633  & ISOGAL-PJ175937.0-290834  &    7.38  &    7.50  &    7.44  &    6.56  &  				\\ 
  681  & ISOGAL-PJ175945.2-290442  &    6.99  &    7.26  &    7.12  &    5.54  &   *C				\\ 
  684  & ISOGAL-PJ175946.1-290318  &    7.03  &    7.44  &    7.24  &    6.97  &   TLE57   Unresolved?    	\\ 
  686  & ISOGAL-PJ175947.0-290227  &    7.27  &          &    7.27  &          &  at edge of raster		\\ 
  689  & ISOGAL-PJ175948.2-290350  &    6.90  &    7.21  &    7.05  &    6.37  &   TLE56			\\ 
  692  & ISOGAL-PJ175949.3-290247  &    7.36  &    7.48  &    7.42  &          &  				\\ 

\end{tabular}                                                       
\end{minipage}
\end{table*}                                                         
 
\section{Correlations with other Catalogues}

The fields observed by ISOGAL overlap completely or in part with areas
surveyed in other ways.

\subsection{Spectroscopic information (NGC\,6522 only)}

The spectroscopic survey of BMB (1984) covers M6 and later M-type giants.
Spatially, more than 80\% of our field is included and almost all their 
stars in the overlap area ($\sim$ 112) were detected. Cross-correlation of 
source positions was performed by means of transparent overlays. The sources 
BMB20 and 21 may both correspond to our no.\ 118. The following BMB 
sources that we do {\it not} detect are of spectral type M6: BMB 41, 56, 
62, 82, 100, and 118. We also did not see BMB\,13 (M5). These are all 
at the early end of the range.

More interesting is the deeper survey of Blanco (1986) which covers a much
smaller portion of the ISOGAL field but includes earlier spectral types,
from M1 onwards. Table 4 summarizes our detections as a function of spectral
type for the part of the field coincident with the deep spectroscopic survey. 
It is clear that our survey cuts off between spectral classes M3 and M4 
(III), where the changeover from majority non-detections to majority 
detections occurs. 

It must also be expected that some brighter objects from the foreground 
will have been included. In fact, there are eight objects in the overlap 
region that were detected in the ISOGAL programme but were not classified as 
having spectra in the range M1--M9 by Blanco (1986). Their $I$-band 
counterparts appear quite bright. They were examined on $I$ and $V$ 
plates (Lloyd Evans, 1976) and were found to be probably non-variable 
and to have colours corresponding to K or early M types. Only one  
was detected at 15\,$\mu$m. The others are almost certainly foreground stars.

\subsection{Photometric information}

\subsubsection{NGC\,6522}

Only one object in the IRAS Point-source Catalog falls within our survey
area. This is IRAS\,17598-2957, which coincides with our no.\ 180, and has
[15] = 5.11 (IRAS 12\,$\mu$m flux = 0.87 Jy). Star 224 has [15] = 4.82 and 
star 12 has [15] = 3.64, so it is surprising that these objects were not 
also detected by IRAS. The cause may be variability or crowding of sources.

A number (56) of our NGC\,6522 stars have been observed by 
Frogel and Whitford (1987) or Tiede, Frogel and Terndrup (1995) on 
the CTIO/CIT near-infrared system. Some of the variables also have 
photometry by Glass and Feast (1982). Fig.\ 7 shows the $(J-K)_0$, 
$K_0$ diagram of the ISOGAL objects that were measured on the CIT/CTIO system.

{\it I}-band photographic photometry was included in the BMB (1984) M-star
survey. Sharples, Walker and Cropper (1990) have pointed out that the stars
with {\it I} mag $<$ 11.8 have a significantly lower velocity dispersion
than fainter ones, indicating that they probably belong to the foreground
disc. A similar effect is seen amongst K giants in the same field 
(Sadler, Terndrup and Rich, 1996). There are 9 BMB stars with {\it I} $<$ 11.8 
amongst the ISO detections, including two Miras (TLE\,238 and TLE\,136) and
a possible semiregular variable (BMB\,18). 

\begin{figure}
\epsfxsize=8cm
\epsffile[28 298 539 786]{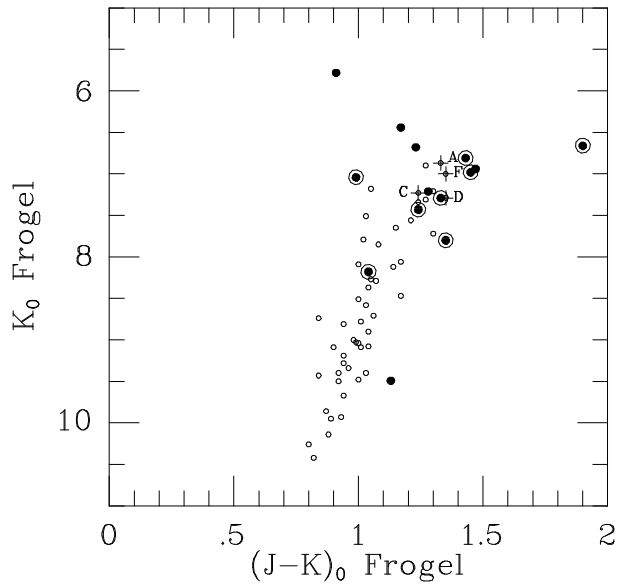}
\caption{$(J-K)_0$, $K_0$ diagram for ISOGAL objects in the NGC\,6522 field 
that have photometry by Frogel and Whitford (1987) or Tiede, Frogel and 
Terndrup (1995) on the CIT/CTIO system. The NGC\,6522 Miras are indicated 
by heavy points. Other high mass-loss objects specially examined (see below, 
section 6) are crossed. Also shown as circled points are Miras in our Sgr 
I field, taken from Glass et al.\ (1995). 
The tip of the RGB is estimated to occur at $K$ $\sim$ 8.2. Note that 
the NGC\,6522 Mira brightest at $K$ (= TLE\,D9) appears unusually blue 
in the Frogel and Whitford photometry. It was considerably redder when 
measured by Glass and Feast (1982).}
\end{figure}

\subsubsection{Sgr I}

There are two IRAS sources within the ISOGAL field, viz no.\ 363 = IRAS
17559-2901 = TLE 53 (IRAS 12\,$\mu$m flux = 1.7 Jy) and no.\ 247 = IRAS 
17558-2858 = TLE 79 (IRAS 12\,$\mu$m flux = 1.19 Jy).

\subsection{Mira-type Variability}

The spectra of Mira variables change by several sub-types around the cycle,
so that a direct correspondence between, for example, period and spectral
sub-type is not to be expected. 

As will be seen, several stars with positions around those of the
less-luminous Miras in the colour-magnitude diagrams have been examined for
large-amplitude variability with negative results. It therefore appears that
our inventory of the Mira variables in these two fields is complete.

\setcounter{table}{3}
\begin{table}
\caption{ISOGAL detections in NGC\,6522 as a function of spectral type 
for stars classified in a deep objective-prism survey by Blanco (1986), 
falling within the area of mutual overlap.}
\begin{tabular}{lll}
M class & Detected & Not detected \\
1       & 0        & 38           \\ 
2       & 5        & 15           \\
3       & 7        & 9           \\
4       & 6        &  5           \\
5       & 6        &  2           \\
6       & 13       &  0           \\
6.5     & 6        &  0           \\
7       & 5        &  0           \\
8       & 1        &  0           \\
\end{tabular}
\end{table}

\subsubsection{NGC\,6522}

Six stars from the ISOGAL field were detected by Lloyd Evans (1976) as 
Mira variables. Four of these are amongst the most luminous at 
15\,$\mu$m, while one, of relatively short period for a ``long-period 
variable" is near the limit of detection. The IRAS source corresponds 
to TLE 228. Table 5 shows the cross-identifications.
The relatively short-period star (115d) TLE\,395 is only just detectable at 
15\,$\mu$m.

\setcounter{table}{4}
\begin{table}
\caption{Stars within the ISOGAL NGC\,6522 field known to be long-period 
variables}

\begin{tabular}{llll}
ISOGAL  &  Lloyd      & Period & Differ- \\
no.\    &  Evans no.\ & days   & ences$^1$ \\
12      & D9      & 400:   & -0.17   \\
180$^2$ & 228     & 270    &  0.00   \\
224     & 403     & 335    &  0.14   \\
245     & 238     & 290:   & -0.18   \\
258     & 395     & 115    &  0.27   \\
491     & 136     & 270    & -0.04    \\
\end{tabular}

Note: A colon denotes uncertainty in the period

$^1$Between the two measurements in LW2. See also table 6.

$^2$ = IRAS\, 17598-2957

\end{table}

The two ISOGAL 7\,$\mu$m exposures were compared by magnitude range (in the
first measurement) and the average differences and their rms values were
found for stars that appear in both. The result is shown in Table 6.

\begin{table}
\caption{Average and r.m.s values of the differences in photometric results 
between the two 7\,$\mu$m exposures of stars in NGC\,6522.}
\begin{tabular}{llll}
Range in      &  No.\ of &  Average    &  RMS        \\
7\,$\mu$m mag &  stars   &  difference &  difference \\
5--6          &  1       &  -0.17      &  0.17       \\
6--7          &  13      &  -0.008     &  0.21       \\
7--8          &  43      &  0.097      &  0.18       \\
8--9          &  113     &  0.111      &  0.22       \\
9--10         &  160     &  0.097      &  0.17       \\
10--11        &  33      &  -.006      &  0.15       \\
\end{tabular}

\end{table}

\subsubsection{Sgr I}

Nine stars within the ISOGAL field are known long-period variables (see
Table 7). These have been followed at {\it JHKL} by Glass et al (1995), who 
list their mean magnitudes and periods.

\begin{table}
\caption{Stars within the ISOGAL Sgr I field known to be long-period 
variables}
\begin{tabular}{llll}
ISOGAL  &  Lloyd     & Period   & Differ-   \\
no.     & Evans no.\ & days$^1$ & ences$^2$ \\
514     & 39         & 336:   & -0.15 \\ 
363$^3$ & 53         & 480    & -0.42 \\
511     & 54         & 293    & -0.38 \\
576     & 55         & 330:   & -0.33 \\
689     & 56         & 235    & -0.31 \\
684     & 57         & 153    & -0.41 \\
78      & 65         & 237    & 0.55  \\
247$^4$ & 79         & 383    & 0.34  \\
402     & 87         & 308    & 0.16  \\
\end{tabular}

Notes:

A colon denotes uncertainty in the period.

$^1$Periods taken from Glass et al (1985) except for ISOGAL no.\ 576 which
is from Lloyd Evans (1976).

$^2$Between the two measurements in LW2. See also table 8.

$^3$ = IRAS17559-2901

$^4$ = IRAS17558-2858

\end{table}

The two ISOGAL 7\,$\mu$m exposures were again compared by magnitude 
range. The result is shown in Table 8.

\begin{table}
\caption{Average and r.m.s values of the differences in photometric results 
between the two 7\,$\mu$m exposures of stars in Sgr I.}
\begin{tabular}{llll}
Range in      &  No.\ of &  Average    &  RMS        \\
7\,$\mu$m mag &  stars   &  difference &  difference \\
4--5          &  1       &   0.06      &  0.06       \\
5--6          &  3       &  -0.010     &  0.34       \\
5--6          &  1$^1$   &  -0.22      &  0.22       \\
6--7          &  16      &  -0.010     &  0.15       \\
6--7          &  11$^1$  &  -0.001     &  0.14       \\
7--8          &  78      &  -0.030     &  0.17       \\
7--8          &  76$^1$  &  -0.020     &  0.14       \\
8--9          &  157     &  -0.020     &  0.17       \\
9--10         &  219     &  -0.017     &  0.17       \\
10--11        &  43      &  -0.12      &  0.24       \\
\end{tabular}

$^1$When known variables are omitted.

\end{table}

\subsubsection{Photometric consistency}

If the known variables are omitted the average and rms differences are 
somewhat reduced. In the case of NGC\,6522, the known variables show only 
small differences and do not materially affect the rms values.

The error in the repeatibility of a single 7\,$\mu$m observation in either 
field is thus about 0.14 mag, where we have divided the RMS difference 
columns in tables 6 and 8 by a factor of $\sqrt{2}$.

\subsection{OH/IR Catalogues}

No known OH/IR sources fall within our fields (of 0.063 square degrees
each). The density of known OH/IR sources in the part of the sky occupied 
by our two fields is only 1--2 per square degree (Sevenster et al., 1997), 
though it is higher at the Galactic Centre ($\sim$10 per square degree for 
a survey of similar depth). The deepest surveys of the Central region 
indicate that the density of OH/IR stars reaches several hundred per 
square degree or about 1/3 the number of known large-amplitude variables 
(Glass et al., 1999). From their $H_K$ colours, the Central region sources 
do not necessarily possess optically thick dust shells at $K$. However, 
it is not clear whether a deeper OH survey of the NGC\,6522 and Sgr I 
fields would yield detections from the objects with moderate dust shells 
reported here. 

\section{The ISOGAL [7]--[15], [15] colour-magnitude diagrams}

Stars detected at both wavelengths are shown in the [7]--[15], [15]
colour-magnitude diagrams (Figs 8 and 9). The fact that these and similar
diagrams discussed below show well-defined sequences implies that they 
are not significantly contaminated by foreground stars.

\begin{figure}
\epsfxsize=8cm
\epsffile[28 298 539 786]{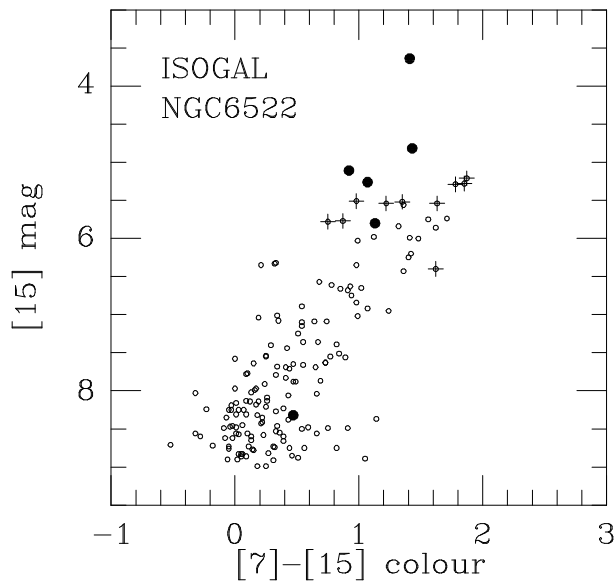}
\caption{ISOCAM colour-magnitude diagram for the 15 $\times$ 15 arcmin$^2$ 
field in the NGC\,6522 Window. Note that many sources are detected only at 
7\,$\mu$m and do not appear here. Known Mira variable stars are indicated 
by heavy points and some other stars chosen for investigation as possible 
high mass-loss objects are crossed. The [7]-[15] colours make use only of 
the second LW2 exposure, which was nearly simultaneous with the LW3.}
\end{figure}

\begin{figure}
\epsfxsize=8cm
\epsffile[28 298 539 786]{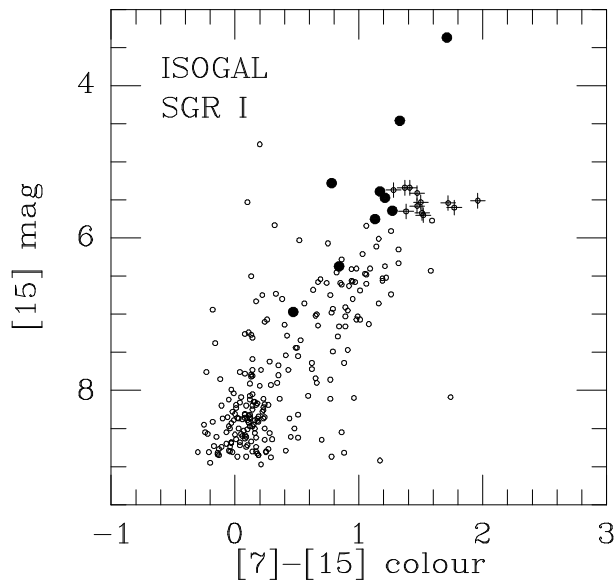}
\caption{ISOCAM colour-magnitude diagram for the 15 $\times$ 15 arcmin$^2$ 
field in the Sgr I Window. See caption of fig.\ 8 for further details.}
\end{figure}

It is instructive to see how the [7]--[15] colours change with spectral 
type, bearing in mind that the few Miras may vary between late types.
This is shown for the NGC\,6522 field in Fig.\ 10, which includes all our 
sources that have been classified by BMB (1984) or Blanco (1986) and have 
been detected in both bands. Note that the fraction of stars 
satisfying both these criteria is almost zero for the early M-types 
and almost one for the later. Except for M sub-types 6, 6.5 and 7, the 
numbers are small. This selection effect arises from the fact that many 
stars fall below our limit in the 15\,$\mu$m band. 

\begin{figure}
\epsfxsize=8cm
\epsffile[28 298 539 786]{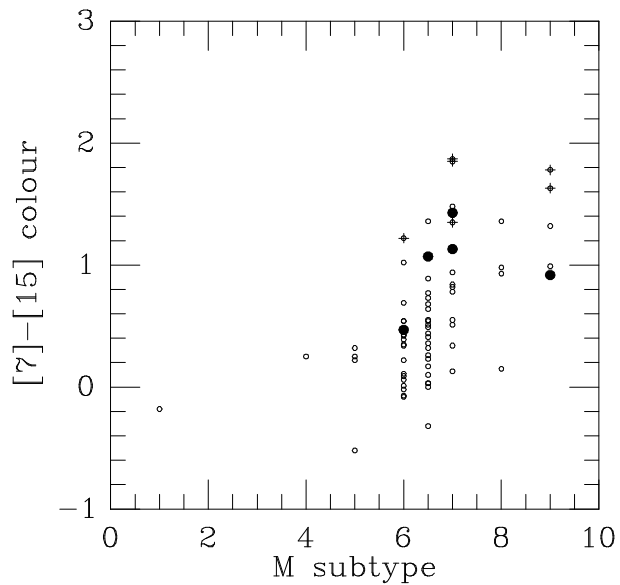}
\caption{[7]--[15] colours vs spectral type for all spectrally classified
stars that were detected at both wavelengths. The heavy dots are known Mira
variables. Other high mass-loss objects are crossed. See text for 
discussion of selection effects. In calculating the colours only the second
LW2 exposure was used.}
\end{figure}

\subsection{Mira [7]--[15] colours}

The [7]--[15] colours of the Mira variables are noticeably displaced from
the extrapolation of the sequence formed by the remainder of the late 
M-stars in the ISOCAM colour-magnitude diagrams (Figs.\ 8 and 9). 
For Miras, the relative flux in the 7\,$\mu$m-band must be greater 
or that in the 15\,$\mu$m-band must be weaker or both. 

O-rich Mira spectra have strong peaks due to the 9.7\,$\mu$m and 18\,$\mu$m 
silicate dust emission features. These are seen clearly in ISO-SWS 
spectra (Onaka et al., 1998). The ISOCAM LW3 band (12--18 $\mu$m) 
is situated longward of the 9.7\,$\mu$m silicate band but is strongly 
influenced for about half of its width by the 18\,$\mu$m feature. So 
far as is known, the LW3 band is not affected significantly by (gaseous) 
molecular absorption, the most conspicuous feature being the CO$_2$ band at 
15.0\,$\mu$m, which appears in some O-rich Miras (Onaka et al, 1997) with
small equivalent width. In many sources an extension of the 10\,$\mu$m 
peak to 13\,$\mu$m, attributed to Al$_2$O$_3$, may also contribute to 
the ISOCAM 15\,$\mu$m band.
 
\section{$K$--[15] colours}

The $K$--[15] colour is primarily a measure of the infrared excess emitted 
by circumstellar dust shells. The $K$-band flux originates in the stellar
photosphere and is only moderately affected ($\leq$ 0.1--0.2 mag overall) 
by CO and H$_2$O absorption bands. As mentioned above, the ISOCAM LW3 
band (12--18 $\mu$m) is strongly influenced for about half of its width 
by the 18\,$\mu$m silicate feature. 

Here we consider those sources detected at 15\,$\mu$m which also have 
ground-based near-infrared observations available from Frogel and 
Whitford (1987). Spectral classifications by BMB (1984) or Blanco (1986)
are also available for these objects. Figure 11 shows their 15\,$\mu$m 
mag vs $K$--[15] colour and Fig.\, 12 shows the 15\,$\mu$m vs M-subtype 
diagram.

\begin{figure}
\epsfxsize=8cm
\epsffile[28 298 539 786]{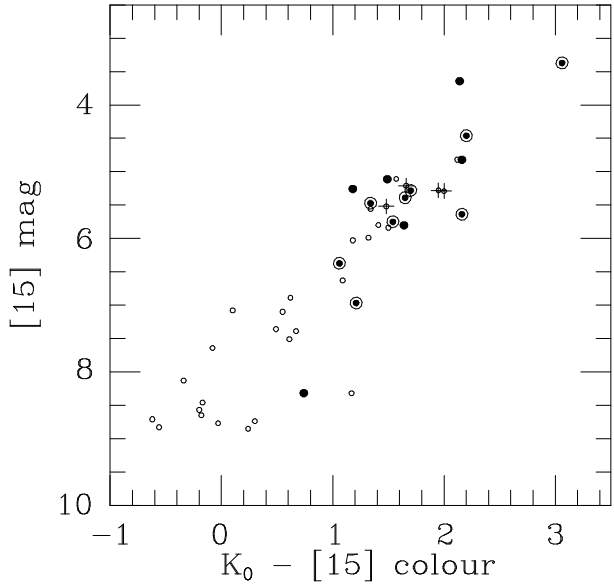}
\caption{15\,$\mu$m mag vs $K_0$--[15] for a subset of stars in the 
NGC\,6522 field detected at 15\,$\mu$m and observed (not simultaneously) 
at $K$ by Frogel and Whitford (1987). Mira variables are indicated by 
heavy points. Other high-mass-loss stars are crossed (see below). For
convenience, the Miras in Sgr I (with $K$ photometry from Glass et al.,
1995) are also plotted (circled points).}
\end{figure}

There is a well-defined sequence of increasing 15\,$\mu$m luminosity with 
$K_0$--[15] colour. Figs 11 and 12, as well as the other C-M diagrams, show
that there is a steady increase of mass-loss from at least the mid-M giants 
to the Miras. It is also clear, for example by adding up the fluxes from
each 15\,$\mu$m magnitude band, that the flux from non-Miras exceeds that 
from the Miras. It is therefore also likely that at least half of the mass 
being returned from the stars to the interstellar medium is from the
non-Miras.

As pointed out by Omont et al (1999c, in preparation), $K_0$ $\sim$ 8.2
corresponds to the tip of the red giant branch (RGB)(Tiede, Frogel and
Terndrup, 1995). This point corresponds (Fig.\ 11) to $K_0$--[15] $\sim$ 0
or [15] $\sim$ 8.2. Thus stars brighter than [15] $\sim$ 8 are on the AGB.

\begin{figure}
\epsfxsize=8cm
\epsffile[28 298 539 786]{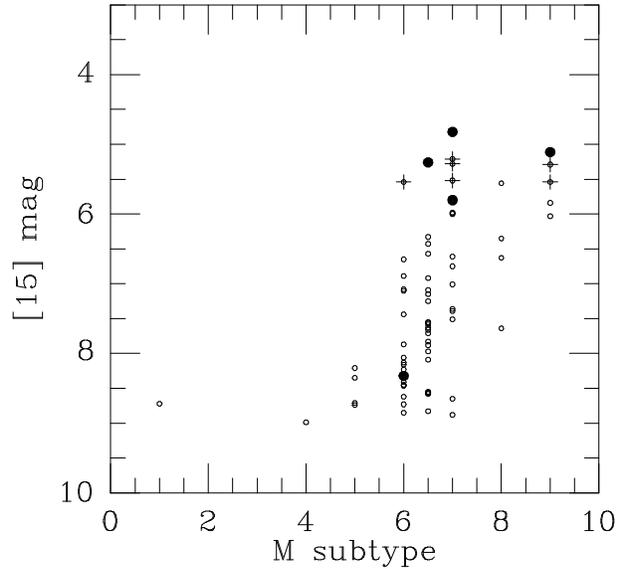}
\caption{15\,$\mu$m mag vs M subtype for a subset of stars in the 
NGC\,6522 field with 15\,$\mu$m detections and spectroscopic information 
available. Symbols as in previous figure.}
\end{figure}

\section{High mass-loss objects with little or no variation}

Five of the known long-period variables in the NGC\,6522 field are near 
the top of the ISOCAM colour-magnitude diagram (Fig.\ 8), but are 
interspersed with other stars that appear similarly luminous and 
somewhat redder. Ten of the latter stars, brighter than [15] = 5.7, 
were selected for detailed investigation. They are denoted by 
letters in Table 2 and are visible as red stars on $V$ and $I$ 
plates (Lloyd Evans, 1976). They have been re-examined for variability. 
Although BMB (1984) assigned very late spectral types and suggested 
variability, none of the ten stars is, in fact, a large-amplitude 
variable. However, variability with small amplitude cannot be excluded 
in about half of them. They therefore represent a type of late M-star 
with high mass loss but small or zero amplitude of variation. The 
colours of four of these stars (A, C, D and F) are shown in 
the $(J-K)_0$, $K_0$ diagram, Fig.\ 7, where they appear to be 
similar to 200--300 day Miras. One of the sample (star B) may be mainly at
maximum, with occasional faint episodes, while stars A and G could be 
variables with occasional bright episodes superimposed on a constant 
background. Such variability in these stars is consistent with 
differences at 7\,$\mu$m between the two observations, especially
for star G (where the change amounts to $\sim$ 0.53 mag).

For Sgr I, as in the case of the NGC\,6522 field, 12 objects with 
15\,$\mu$m mags similar to the known Miras and somewhat redder 
colours were checked for variability by Lloyd Evans on the plate 
material. In all cases, these objects could be identified with 
very red stars. Only two were found to be fairly certain variables 
of low amplitude (F and H). Both these stars spend most of their 
time at maximum, with two fainter episodes separated by about 250 
days for F and a single one for H. 

The positions of these objects in the [7]--[15], [15] diagram (Figs. 8, 9 )
suggest that they form a continuation of the general sequence of 
late M stars and are not similar to the Mira variables. Unfortunately, 
only four of them, BMB\,28, BMB\,46, BMB\,86 and BMB\,179, have been 
measured in the near-infrared. Their $K_0$--[7] indicies are near-zero
(Fig.\ 14).

Two of these non-Mira high mass-loss objects have been classified
spectroscopically by BMB (1984) as M9, a spectral type that is 
usually associated with large-amplitude variability. It is known 
that the onset of Mira-type behaviour occurs at later types with 
increasing metallicity.
The existence of this class of stars may therefore be a consequence of
super metal richness. Alternatively, variability may have ceased temporarily
but the dust shells have not yet dissipated.

Sloan and Price (1995) and Sloan, LeVan and Little-Marenin (1996) have 
shown that certain irregular and semi-regular AGB variables have dust 
excesses in IRAS spectra and are associated with the appearance of the 
13\,$\mu$m dust feature. By obtaining mid-infrared spectra for our objects 
or by obtaining more precise variability information it will be possible 
to decide if they are identifiable with this category. They may also be
similar to the `red' O-rich SRb variables of Kerschbaum and Hron (1996).

\section{The $K_0$--[7] colours}

The 56 objects of NGC\,6522 with $JHK$ photometry and spectral 
classifications are plotted in figure 13, which shows their $K_0$--[7] 
colours and spectral types.

\begin{figure}
\epsfxsize=8cm
\epsffile[28 295 539 786]{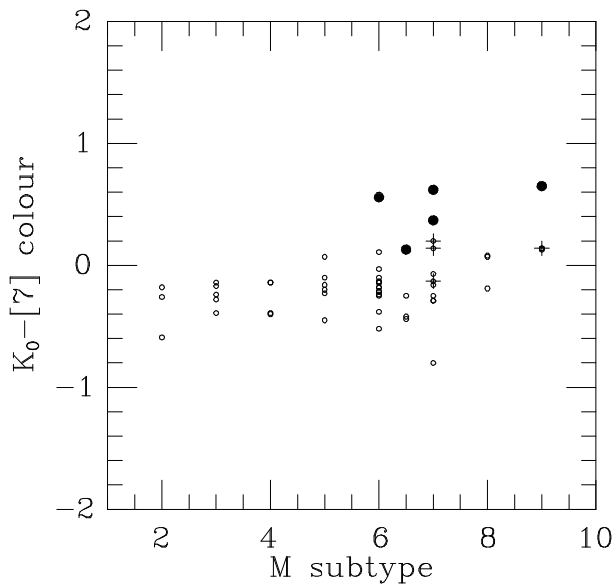}
\caption{$K_0$--[7] colours vs spectral type for NGC\,6522 field stars 
that have $K_0$ photometry by Frogel and Whitford (1987) 
and spectral classifications by BMB (1984) or Blanco (1986).
This colour index is predominantly negative for unreddened M giants, 
except for some of the latest spectral types. Note the $K$ and 7\,$\mu$m
photometry was not simultaneous, and that averaged 7\,$\mu$m mags were used
when available. Miras are shown as heavy dots and other high mass-loss 
objects are crossed.}
\end{figure}

These colours are unexpectedly negative in most cases, especially 
for the earlier M-subtypes. While part of this could be due to the 
small uncertainty in the absolute 7\,$\mu$m photometry in crowded fields, 
ISO SWS spectra of late-type O-rich M stars frequently show a broad 
absorption shortward of the strong silicate dust peak at 10$\mu$m. 
This region is known to be affected by the SiO fundamental at 
$\sim$ 7.9\,$\mu$m (Cohen et al., 1995). Carefully calibrated 
spectrophotometry of $\beta$ Peg (M2.5\,II--III) shows absorption 
at around 6\,$\mu$m attributed to the $\nu_2$ band of water vapour 
centred at 6.25\,$\mu$m. Spectra of O-rich M-type giants that also 
demonstrate these features are presented by Tsuji et al.\ (1998), 
who envisage that they arise in a warm absorbing layer somewhat 
above the photosphere. Approximate calculations based on the Tsuji 
et al.\ spectra yielded $K_S$--[7] $\sim$ --0.1 for $\beta$
Peg but slightly positive $K_{\bf S}$--[7] for SW Vir 
(M7III)\footnote{Spectrophotometry shows that $K$ -- $K_{\it DENIS}$ 
should have values of 0.02 to 0.07 for Miras because the $K_S$ band 
does not include the first overtone band of CO.}. The presence of 
significant amounts of dust could be an influence in the latter case.

The depth of the CO fundamental band in late M-type giants is known to 
have a similar effect on the $L-M$ colours, which tend to have low or 
negative values.

Figure 13 shows that in the latest M-type giants the $K_0$--[7] colour 
approaches zero or may even become positive. It is known that the overtone SiO 
absorption band at 3.95--4.1\,$\mu$m is strong in semiregular variables 
but in Miras its strength varies with time and it can become weak 
(Aringer et al., 1995; Rinsland and Wing, 1982). 

The optical depth of the circumstellar dust in the stars that we are
discussing is very low in the earlier spectral types, rising somewhat
towards the Miras but never becoming high. The emission at 9.7 and 
18\,$\mu$m is probably dominated by particles at a fairly uniform 
temperature around 1000\,K, where condensation of silicate dust grains 
first becomes possible in the stellar wind. Thus the 15\,$\mu$m fluxes 
are largely a measure of the mass of grains present and ultimately 
of the mass-loss rate from the star. The near-infrared $K$ flux   
will not be affected by {\it emission} from the dust but will be 
reduced slightly by the {\it absorption} it causes at shorter wavelengths.
The opacity of silicate dust at 7\,$\mu$m is very low (see e.g. Schutte and
Tielens, 1989) so that the trend of $K$--[7] colour with spectral 
type that we observe could partly be caused by increasing extinction at 
$K$ and decreasing SiO and possibly H$_2$O molecular band strengths 
at 5.5--8\,$\mu$m. However, dust emission at 7\,$\mu$m is certainly
responsible for the largest values observed in LPVs, as shown by 
Groenewegen (private communication) for various dust models.

\subsection{The $K_0$--[7], [7] colour-magnitude diagram}

In the  $K_0$--[7], [7] colour-magnitude diagram (Fig.\ 14) we include 
the Miras from both fields. It is seen that almost all of the Miras 
are brighter and redder at 7\,$\mu$m than the other red giants. A diagram 
involving the $K_0$--[7] colour will therefore be the best criterion for 
detection of large-amplitude LPVs. 

\begin{figure}
\epsfxsize=8cm
\epsffile[28 298 539 786]{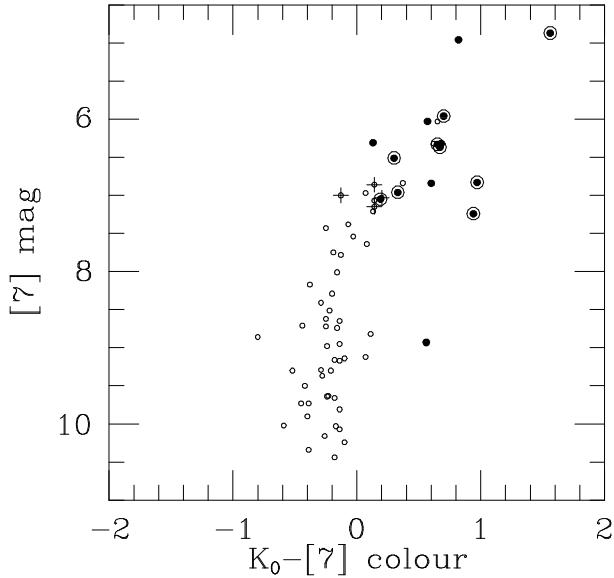}
\caption{The $K_0$--[7], [7] colour-magnitude diagram, including Miras from
NGC\,6522 (heavy dots) and Sgr I (circled). The Miras are almost completely 
distinguishable from other stars by their bright 7\,$\mu$m mags. Note that 
the $K$ and 7\,$\mu$m photometry was not simultaneous and that averaged 
7\,$\mu$m mags were used when available. Crosses denote other high
mass-losing objects.}
\end{figure}

\section{Absence of very luminous sources}

Let us recall that, for the class of sources that we detect, the $K$ 
bolometric magnitude correction is practically constant ($\sim$ 3.0; 
Groenewegen, 1997). With a distance to the Galactic Centre of 8.5\,kpc, 
$M_{\rm Bol}$ $\sim$ $K_0$ - 11.65. The brightest stars in our fields 
therefore have bolometric magnitudes $>$ --5.9. Neither field includes 
any luminous dust-enshrouded AGB sources of the type found in the Large 
Magellanic Cloud by Wood et al.\ (1992), which reach luminosities of 
$M_{\rm Bol}$ $\sim$ --7.5.  

\section{Mid-infrared period-colour relation for Miras}

The Mira variables show a clear period-colour relation with very moderate 
scatter in the mid-infrared (Fig.\ 15). The [7] and [15] photometry is 
almost simultaneous, so that such scatter as exists is not attributable to 
variation between measurements. Following the discussion in section 7, 
this relation confirms that the mass-loss rate in Miras is
directly related to their periods (see e.g., Whitelock et al., 1994).

The colour-magnitude diagrams for each of our two fields do not appear to
contain any stars brighter than the known Miras. It is therefore likely that
the census of stars at the long-period end of the the range is complete and
that there is no evidence for long-period variables in our fields with 
periods longer than those already known, i.e. $\sim$ 700 d (see Glass et 
al., 1995).

\begin{figure}
\epsfxsize=8cm
\epsffile[28 298 539 786]{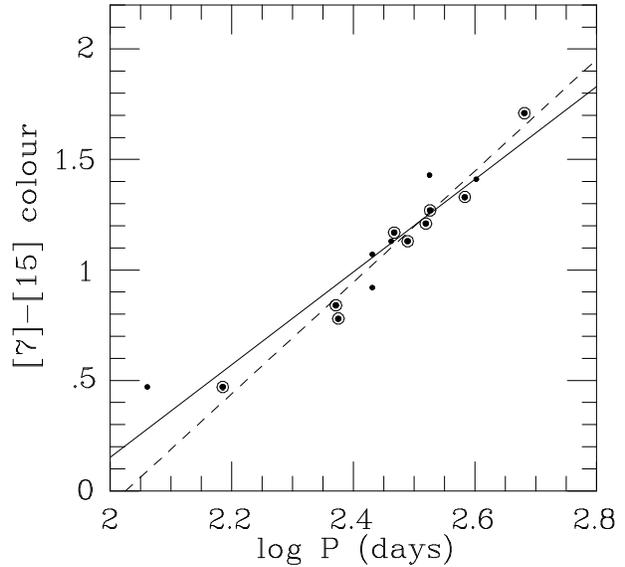}
\caption{Mid-infrared period-colour relation for Mira variables, based on
the Miras in the NGC\,6522 (points) and Sgr I fields (circled points). 
The full regression line (solid) follows the equation [7]--[15] = 2.10 log 
$P$ -- 4.05. The rms differences between the fit and the observed points 
is 0.12 mag, rather better than expected from the photometric errors. 
If the shortest-period point is omitted, the fit (dashed line) becomes 
[7]--[15] = 2.52 log $P$ -- 5.11, with rms difference = 0.07. Only the 
second 7\,$\mu$m exposure was used in forming the colours.}
\end{figure}

\section{Star Counts}

Determination of the spatial density distributions of the various
stellar populations in the central Galaxy, and especially the bulge,
is one of the key scientific goals of the ISOGAL project. A detailed
analysis, covering several fields, and with careful consideration of
completeness, extinction, etc, is in preparation. For the present we
consider simply the differential number counts between the two fields
at magnitudes above the completeness limit. At [15] this covers the
AGB above the RGB-tip, while at [7] the limit extends approximately
one mag below the RGB tip, to [7]$\sim 9.5$, using $K$ -- [15] and 
$K$ -- [7] colours from Figs 11 and 14 respectively. 

The ratio of the surface density of sources is, within the sampling
errors, identical for both RGB and AGB stars, indicating that there
are no steep population gradients apparent. It is also identical at
both [7] and [15] microns. The surface density ratio
is $$\rm  \rho(Sgr\;I)/\rho(NGC\,6522) = 1.85.$$

This count ratio corresponds to an exponential scale height of
2.0$^{\circ}$, or $\sim 280$\,pc. That is, the inner bulge minor axis scale
height is the same as that of the Galactic disk.

A more interesting comparison is with the scale height derived from
analysis of the COBE/DIRBE surface brightness observations of the
inner Galaxy. These have been analysed most completely by Binney,
Gerhard, and Spergel (1997), whose best-fit model has a bulge density
distribution which is a truncated power-law:
$$f_b = f_0{{{\rm e}^{-a^2/a_m^2}}\over{(1+a/a_0)^{1.8}}}$$
with $a_m$=1.9kpc, $a_0$=100pc, $f_0$ a normalisation constant, and $a
= (z/\xi)$ for ($x = y = 0$), where $z$ is the minor axis distance 
and $\xi$ an axial ratio, with $\xi = 0.6$ best fitting the data. 
This density profile, which interpolates between the central luminosity
spike, which follows an $R^{-1.8}$ luminosity profile, and the outer
bulge, which follows an $R^{-3.7}$ profile, is also used in the models
of Kent, Dame and Fazio (1991) and the kinematic analysis 
of Ibata and Gilmore (1995). It is known to provide an acceptable 
description of the bulge from latitudes of 4$^{\circ}$ to at least
12$^{\circ}$. It remains to be tested at intermediate latitudes.

Direct comparison of the ISOGAL data
with this function is of specific interest, since the fields surveyed
here are included in the COBE/DIRBE analysis, and are of low
reddening. The COBE/DIRBE data are however integrated light, and so
must have a statistical correction for the foreground disk. This can
be a complex function of wavelength and spatial resolution, especially
at low latitudes (see Unavane et al., 1998, and Unavane and Gilmore,
1998) for a  more complete description). Our ISOGAL source counts are
however strongly biased against the foreground disk, being
dominated by sources in the central Galaxy.
Thus, the ISOGAL observations test directly the analysis of the
integrated light. Our present analysis however does not consider line
of sight depth effects in the bulge itself. 

With the adopted vertical axis ratio $\xi=0.6$ the predicted count
ratio in the ISOGAL data is $\rm  \rho(Sgr\;I)/\rho(NGC\,6522) = 2.04$,
somewhat larger than the observed value. Interestingly, near-exact agreement
with the model above follows with an axis ratio $\xi =1.0$, which
predicts $\rm  \rho(Sgr\;I)/\rho(NGC\,6522) = 1.81$. This 
systematic discrepancy between the model and our data is in agreement
with the small systematic residuals emphasised to exist by Binney
et al.\ (1997) and shown in their figure 2. 

\section{Conclusions}

Most of the detected objects are late-type M stars, with a cut-off for those
earlier than about M3--M4.

There is a continuous sequence of increasingly mass-losing objects from the
mid-M-type giants to the long-period Miras, which are the most luminous 
stars in the field.

There appears to be no component of dust-enshrouded very long-period OH/IR
stars or similar objects faint even at $K$ in these fields. The upper limit
of luminosities remains at $M_{\rm Bol}$ $\sim$ --5.7  and the upper limit 
of periods remains at about 700 days, as determined from near-infrared 
studies (Glass et al., 1995).

There is a group of late-type M-stars on the AGB which are not 
large-amplitude variables but may be irregular or of small amplitude, 
that have luminosities similar to Mira variables in the 200--300 day period
range and show redder [7]--[15] but bluer $K_0$--[7] colours.

The ISOCAM 7\,$\mu$m band is almost certainly affected by molecular 
absorption in ordinary M-giant stars. However, Mira variables are brighter 
than other stars possibly in part because of reduced SiO and H$_2$O 
absorption.

The results from these fields should form a template for analyses of more
heavily obscured regions about which little is known from visible-light
studies.

\section{Acknowlegments}

We would like to acknowledge the Les Houches 1998 summer school 
for access  to the JUN98 version of CIA. We acknowledge useful 
discussions with K.S. Krishnaswamy, TIFR.  Eric Copet is thanked for 
his help with the Unix scripts and Martin Groenewegen for his comments on
dust at 7\,$\mu$m. ISG wishes to acknowledge the hospitality 
of IAP during part of this work. SG and MS acknowlege receipts of 
fellowships from the Ministere des Affaires Etrang\`{e}res, France, 
and ESA respectively.


\begin{thebibliography}{}

\bibitem{} Abergel A., Miville-Desch\^enes M.A., D\'esert F.X., P\'erault
 M., Aussel H., Sauvage M.\ 1998 \\
 http://www.iso.vilspa.esa.es/users/expl\_lib/CAM\_list.html
\bibitem{} Alard et al.\ 1999, in preparation
\bibitem{} Aringer B., Wiedemann G., K\"{a}ufl H.U., Hron J.\ 1995, Ap Sp
 Sci, 224, 421
\bibitem{} Binney J.J., Gerhard O.E., Spergel D.N.\ 1997, MNRAS 288, 365
\bibitem{} Blanco V.M.\ 1986, AJ, 91, 290
\bibitem{} Blanco V.M., McCarthy M.F., Blanco B.M.\ 1984, AJ, 89, 636 
 (BMB)
\bibitem{} Blommaert J.A.D.L.\ 1998, {\it ISOCAM Photometry Report},\\
http://www.iso.vilspa.esa.es/users/expl\_lib/CAM\_list.html
\bibitem{} Cesarsky C.J. et al.\ 1996, A\&A, 315, L32
\bibitem{} Cohen M., Witteborn F.C., Walker R.G., Bregman J.D., Wooden 
 D.H.\ 1995, AJ 112, 241
\bibitem{} Feast M.W.\ 1985, {\it Light on Dark Matter}, Israel F.P., Reidel, 
 Dordrecht, p.\ 339.
\bibitem{} Frogel J.A., Whitford A.E.\ 1987, ApJ, 320, 199
\bibitem{} Ganesh S. et al.\ 1999, in preparation
\bibitem{} Glass I.S.\ 1986, MNRAS 221 879
\bibitem{} Glass I.S.\ 1993, {\it IAU Symp.\ 153: Galactic Bulges},  
 Habing H., Dejonge H., Kluver, Dordrecht, p.\ 21
\bibitem{} Glass I.S., Feast M.W.\ 1982 MNRAS, 198, 199
\bibitem{} Glass I.S., Matsumoto S., Carter B.S., Sekiguchi K.\ 1999 {\it IAU 
 Symp.\ 191, Asymptotic Giant Branch Stars}, Le Bertre T., L\`{e}bre A.,
 Waelkens C., PASP Conf Ser, in press 
\bibitem{} Glass I.S., Whitelock P.A., Catchpole R.M., Feast M.W.\ 
 1995 MNRAS, 273, 383
\bibitem{} Groenewegen, M.\ 1997, in {\it The Impact of Large-Scale Near-IR
 Surveys}, Garzon F. et al., Kluwer, p.\ 165
\bibitem{} Ibata R.A., Gilmore G.\ 1995, MNRAS 275, 605
\bibitem{} Kent S.M., Dame T., Fazio G.\ 1991, ApJ 378, 131
\bibitem{} Kerschbaum R., Hron J.\ 1996, A\&A, 308, 489 
\bibitem{} Lloyd Evans T.\ 1976, MNRAS, 174, 169 (TLE)
\bibitem{} Ojha D., Omont A., Simon G.\ 1997, {\it IAU Symposium 184: 
 The Central Regions of the Galaxy and Galaxies}, Sofue Y., p.\ 43 
\bibitem{} Omont A. and the ISOGAL Collaboration 1999a, to appear in 
 {\it Astrophysics with Infrared Surveys: A Prelude to SIRTF}, 
 Proc. First SIRTF Conf., Bicay M., PASP Conference Series
\bibitem{} Omont A. et al.\ 1999b, to appear in {\it The Universe as seen 
 by ISO}, Cox P., Kessler M.F., ESA Special Publications Ser. (SP-427)
\bibitem{} Omont A. et al.\ 1999c, in preparation  
\bibitem{} Onaka T, Yamamura I., de Jong T., Tanab\'{e} T., Hashimoto O.,
 Izumiura W.\ 1998, Ap Sp Sci, 255, 331
\bibitem{} Onaka T., de Jong T., Yamamura I., Tanab\'{e} T., Hashimoto O.,
 Izumiura W.\ 1997, Proc. First ISO Workshop on Analytical Spectroscopy,
 ESA-SP-419 
\bibitem{} P\'{e}rault M., Omont A. et al.\ 1996, A\&A, 315, L165
\bibitem{} Rinsland C.P., Wing R.F.\ 1982, ApJ, 262, 201
\bibitem{} Sadler E.M., Rich R.M., Terndrup D.M.\ 1996, AJ, 112, 171
\bibitem{} Schutte W.A., Tielens A.G.G.M.\ 1989, ApJ, 343 369 
\bibitem{} Sevenster M.N., Chapman J.M., Habing H.J., Killeen N.E.B.,
 Lindqvist M.\ 1997, A\&AS, 122, 79.
\bibitem{} Sharples R., Walker A., Cropper A.\ 1990, MNRAS, 246, 54 
\bibitem{} Sloan G.C., LeVan P.D., Little-Marenin I.R.\ 1996, ApJ 463, 310
\bibitem{} Sloan G.C., Price S.D.\ 1995, ApJ 451, 758
\bibitem{} Stanek K.Z.\ 1996, ApJ 460, L37.
\bibitem{} Tiede G.P, Frogel J.A., Terndrup D.M.\ 1995, AJ, 110, 2788
\bibitem{} Tsuji T., Ohnaka K., Aoki W., Yamamura I.\ 1998, Ap Sp Sci 
 255, 293
\bibitem{} Unavane M., Gilmore G.\ 1998, MNRAS 295, 145
\bibitem{} Unavane M., Gilmore G., Epchtein N., Simon G., Tiphene D., deBatz
 B.\ 1998, MNRAS 295, 119
\bibitem{} Whitelock P., Menzies J., Feast M., Catchpole R., Marang F.,
Carter B.\ 1994 MNRAS 276 219
\bibitem{} Wood P.R., Whiteoak J.B., Hughes M.G., Bessell M.S., Gardner F.F.
 Hyland A.R.\ 1992, ApJ 397, 552
\end{thebibliography}
\end{document}